\documentclass[twocolumn,aps,prl,showpacs,superscriptaddress]{revtex4}

\usepackage{graphicx}
\usepackage{amsmath,amssymb}
\usepackage{color}
\def\endproof{\vrule height6pt width6pt depth0pt} 

\begin{document}

%%%%%%%%%%%%%%%%%%%%%%%%%%%%%%%%%%%%%%%%%%%%%%%%%%%%%%%%%%%%%%%%%%%

\title{Twin inequality for fully contextual quantum correlations}

%%%%%%%%%%%%%%%%%%%%%%%%%%%%%%%%%%%%%%%%%%%%%%%%%%%%%%%%%%%%%%%%%%%

\author{Ad\'an Cabello}
%\email{adan@us.es}
\affiliation{Departamento de F\'{\i}sica Aplicada II, Universidad de Sevilla, E-41012 Sevilla, Spain}

%%%%%%%%%%%%%%%%%%%%%%%%%%%%%%%%%%%%%%%%%%%%%%%%%%%%%%%%%%%%%%%%%%%

\date{\today}

%Version 1: July 28, 2012 (In the plane between Sevilla and Roma).
%Version 5: September 1, 2012 (Sevilla). The first version in the arXiv.
%Version 6: September 11, 2012 (Sevilla). The first version in PRL.
%Version 7: September 11, 2012 (Sevilla). Simple proof of why requires dimension 6.
%Version 9: January 16, 2013 (Concepcion).
%Version 10: January 26, 2013 (Belo Horizonte). After PRA proofs.

%%%%%%%%%%%%%%%%%%%%%%%%%%%%%%%%%%%%%%%%%%%%%%%%%%%%%%%%%%%%%%%%%%%

\begin{abstract}
Quantum mechanics exhibits a very peculiar form of contextuality. Identifying and connecting the simplest scenarios in which more general theories can or cannot be more contextual than quantum mechanics is a fundamental step in the quest for the principle that singles out quantum contextuality. The former scenario corresponds to the Klyachko-Can-Binicio\u{g}lu-Shumovsky (KCBS) inequality. Here we show that there is a simple tight inequality, twin to the KCBS, for which quantum contextuality cannot be outperformed. In a sense, this twin inequality is the simplest tool for recognizing fully contextual quantum correlations.
\end{abstract}

%%%%%%%%%%%%%%%%%%%%%%%%%%%%%%%%%%%%%%%%%%%%%%%%%%%%%%%%%%%%%%%%%%%

%03.65.Ta: Foundations of quantum mechanics; measurement theory
%03.65.Ud: Entanglement and quantum nonlocality
%(e.g. EPR paradox, Bell's inequalities, GHZ states, etc.)
%42.50.Xa: Optical tests of quantum theory
%02.10.Ox: Graph theory

\maketitle

%%%%%%%%%%%%%%%%%%%%%%%%%%%%%%%%%%%%%%%%%%%%%%%%%%%%%%%%%%%%%%%%%%%

{\em Introduction.} For more than 50 years, we have known that quantum correlations cannot be reproduced with noncontextual hidden variable (NCHV) theories in which the results of measurements are independent of other compatible measurements \cite{Specker60, Bell66, KS67} (two measurements are compatible when, for any preparation, each measurement always gives the same outcome, no matter how many times the measurements are performed or in which order).
Recent research has taken a step forward, pointing out that quantum mechanics (QM) exhibits a very peculiar form of contextuality \cite{CSW10} and conjecturing that the physical principle responsible of this form of contextuality may be the physical principle from which the whole QM derives \cite{Fuchs11,Cabello11,Cabello12}. This approach has started by exploring the limits of quantum contextuality in contrast to those of more general theories \cite{CSW10}, trying to understand the reasons that single out quantum contextuality among more general forms of contextuality. Within this program, it is of fundamental importance to identify the simplest scenario in which more general theories cannot be more contextual than quantum mechanics, and connect it with the simplest scenario in which they can be more contextual.

The standard method to recognize contextual correlations is through the violation of noncontextuality (NC) inequalities, which are inequalities involving joint probabilities of compatible measurements on the same system, and which are satisfied by any NCHV theory and violated by QM. The simplest physical system exhibiting quantum contextuality for repeatable measurements is a three-level quantum system \cite{Specker60, Bell66, KS67}. Two-level systems can only show some forms of contextuality when generalized measurements are considered \cite{Cabello03,Spekkens05}. The simplest NC inequality violated by a three-level quantum system is the Klyachko-Can-Binicio\u{g}lu-Shumovsky (KCBS) inequality \cite{KCBS08}, which requires five experiments, each of them involving two compatible yes-no tests represented in QM by projectors $\Pi_i=|v_i\rangle \langle v_i|$ onto unit vectors $|v_i\rangle$, with possible results $1$ (yes) and $0$ (no). The KCBS inequality is tight (i.e., is a facet of the corresponding polytope of noncontextual correlations) and can be written as
\begin{equation}
 \label{KCBS}
 \frac{1}{2} \sum_{i=0}^4 P(\Pi_i+\Pi_{i+1}=1) \stackrel{\mbox{\tiny{ NCHV}}}{\leq} 2 \stackrel{\mbox{\tiny{ QM}}}{\leq} \sqrt{5} \stackrel{\mbox{\tiny{ GP}}}{\leq} \frac{5}{2},
\end{equation}
where the sum in the subindexes is defined modulo 5 (i.e., $4+1=0$), $P(\Pi_i+\Pi_{i+1}=1)$ denotes the probability that exactly one of $\Pi_i$ and $\Pi_{i+1}$ has the result 1 [i.e., $P(\Pi_i+\Pi_{i+1}=1):=P(\Pi_i=1 \;{\rm XOR}\; \Pi_{i+1}=1)$], $\stackrel{\mbox{\tiny{ NCHV}}}{\leq} 2$ indicates that $2$ is the maximum for NCHV theories, $\stackrel{\mbox{\tiny{ QM}}}{\leq} \sqrt{5}$ indicates that $\sqrt{5} \approx 2.236$ is the maximum for QM (using quantum systems of arbitrary dimension), and $\stackrel{\mbox{\tiny{ GP}}}{\leq} \frac{5}{2}$ indicates that $\frac{5}{2}$ is the maximum for general probabilistic (GP) theories, defined as those satisfying that the sum of the probabilities of mutually exclusive events cannot be higher than one.

Notice the existence of a gap in \eqref{KCBS} between the maximum for QM and the maximum for GP theories. This means that, for the KCBS inequality, more general theories beyond QM can exhibit correlations which are more contextual than those in QM.

Recent research has identified NC inequalities for which the maximum quantum violation saturates the maximum for GP theories. These inequalities reveal fully contextual quantum correlations, defined as those that {\em cannot} be expressed as a nontrivial convex sum of noncontextual and contextual correlations or, equivalently, those having a zero noncontextual content \cite{ADLPBC12}.

A fundamental open question is which is the simplest NC inequality capable of revealing fully contextual quantum correlations. Here we show that the one requiring fewer experiments involving yes-no tests $\Pi_i$ is the following one:
\begin{equation}
 \label{C12}
 \frac{1}{2} \sum_{i=0}^4 P(\Pi_i+\Pi_{i+1}+\Pi_{i+5}+\Pi_{i+7}=1) \stackrel{\mbox{\tiny{ NCHV}}}{\leq} 2 \stackrel{\mbox{\tiny{QM, GP}}}{\leq} \frac{5}{2},
\end{equation}
where the sum in the subindexes is defined such that $4+1=0$ and $3+7=5$. As the KCBS inequality, testing inequality \eqref{C12} requires only five experiments involving yes-no tests $\Pi_i$ (but now $i=0,\ldots,9$). As the KCBS inequality, inequality \eqref{C12} is tight (this can be checked, e.g., using \texttt{porta} \cite{Porta}). An interesting property of inequality \eqref{C12} is that its quantum violation requires quantum systems of dimension six (or higher). To our knowledge, this is the first time that quantum systems of dimension six are needed for the violation of a fundamental NC (or Bell) inequality.

The rest of this Rapid Communication is dedicated to prove all these statements, introduce the quantum state and yes-no tests needed to reveal correlations with zero noncontextual content through the violation of inequality \eqref{C12}, and briefly discuss how to observe them experimentally.

%%%%%%%%%%%%%%%%%%%%%%%%%%%%%%%%%%%%%%%%%%%%%%%%%%%%%%%%%%%%%%%%%%%
% Fig. 1
%%%%%%%%%%%%%%%%%%%%%%%%%%%%%%%%%%%%%%%%%%%%%%%%%%%%%%%%%%%%%%%%%%%

\begin{figure}[t]
\centerline{\includegraphics[width=0.80\columnwidth]{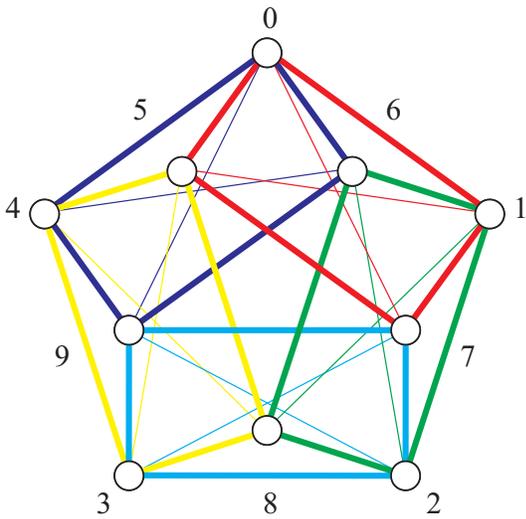}}
\caption{(Color online) Graph representing the orthogonalities between the vectors $|v_i\rangle$ used to construct the ten yes-no tests $\Pi_i$ in inequality \eqref{C12}. Each vertex represents a vector and two vertices are adjacent if and only if they are orthogonal. The five sets of four mutually orthogonal vectors are distinguished using five different colors. Each of these sets corresponds to an experiment for testing inequality \eqref{C12}.}\label{Fig1}
\end{figure}

%%%%%%%%%%%%%%%%%%%%%%%%%%%%%%%%%%%%%%%%%%%%%%%%%%%%%%%%%%%%%%%%%%%

{\em Theorem 1.} Inequality \eqref{C12} is the NC inequality containing the smallest number of sets of compatible yes-no tests $\Pi_i$ and capable of revealing fully contextual quantum correlations.

{\em Proof. } Any linear combination of joint probabilities of compatible measurements of observables $\Pi_i$, such as those appearing in any NC inequality, can be expressed as
\begin{equation}
 \beta = \sum_{i \in V(G)} \sum_{(j,\ldots,k) \in C(i)} w_i^{(j,\ldots,k)} P(\Pi_i=1, \Pi_j=0,\ldots, \Pi_k=0),
\end{equation}
where $V(G)$ is the set of vertices of the graph $G$ in which every $\Pi_i$ is represented by a vertex and two vertices are adjacent if and only if the corresponding projectors are orthogonal, $C(i)$ is the set of cliques (i.e., complete subgraphs of $G$) of an edge clique cover of $G$ (i.e., a set of cliques in $G$ that together cover all of the edges) containing $i$; $w_i^{(j,\ldots,k)} \ge 0$, and $\sum_{(j,\ldots,k) \in C(i)} w_i^{(j,\ldots,k)}=1$. Since $\Pi_i$, $\Pi_j$,\ldots, $\Pi_k$ are compatible, $P(\Pi_i=1, \Pi_j=0,\ldots, \Pi_k=0)=P(\Pi_i=1)$. Therefore,
\begin{equation}
 \beta = \sum_{i \in V(G)} P(\Pi_i=1).
\end{equation}

It has been shown \cite{CSW10} that the maximum of $\beta$ for NCHV theories, QM, and GP theories are given by three characteristic numbers of $G$.
Specifically, the maximum of $\beta$ for NCHV theories is given by the independence number of $G$, $\alpha(G)$ \cite{Diestel10}, which is the maximum number of pairwise nonadjacent vertices in $G$. The maximum of $\beta$ for QM is given by the Lov\'asz number of $G$, $\vartheta(G)$ \cite{Lovasz79}, which is
\begin{equation}
 \label{lovasz}
 \vartheta(G) = \max \sum_{i \in V(G)} |\langle\psi|v_{i}\rangle|^{2},
\end{equation}
where the maximum is taken over all unit vectors $|\psi\rangle$ and $|v_{i}\rangle$ in any dimension, where each $|v_{i}\rangle$ corresponds to a vertex of $G$, and two vertices are adjacent if and only if the vectors are orthogonal. Therefore, the set $\{|v_{i}\rangle\}$ provides an orthogonal representation of $G$ (i.e., it allows us to assign one vector to each vertex in $G$ such that adjacent vertices correspond to orthogonal vectors). Finally, the maximum of $\beta$ for GP theories is given by the fractional packing number of $G$, $\alpha^*(G)$ \cite{SU97}, defined as
\begin{equation}
 \alpha^*(G) = \max \sum_{i\in V(G)} w_i,
\end{equation}
where the maximum is taken for all $0 \leq w_i\leq 1$ and for all cliques $c_j$ of $G$, under the restriction $\sum_{i \in c_j} w_i \leq 1$.

For example, the graph $G$ associated with the KCBS inequality is a pentagon, which has $\alpha(G)=2$, $\vartheta(G)=\sqrt{5}$, and $\alpha^*(G)=\frac{5}{2}$, which are the three bounds in inequality \eqref{KCBS}.

From those results follows that a necessary and sufficient condition for a NC inequality to reveal fully contextual quantum correlations is to be associated with a graph $G$ such that $\alpha(G)<\vartheta(G)=\alpha^*(G)$. These three numbers have been calculated for all graphs with less than 11 vertices \cite{CDLP11} and, among them, there are only four graphs with this property, all of them with ten vertices.

The NC inequality requiring the smallest number of experiments involving yes-no tests $\Pi_i$ corresponds to the graph $G$ with the smallest edge clique cover number $\theta'(G)$, which is the cardinal of an optimal edge clique cover of $G$ (i.e., one with the smallest number of cliques). $\theta'(G)$ gives the minimum number of experiments needed to test the NC inequality. Among the four ten-vertex graphs with $\alpha(G)<\vartheta(G)=\alpha^*(G)$, the one with the smallest $\theta'(G)$ is the one in Fig.~\ref{Fig1}. A simple inspection shows that this graph is the one associated with inequality \eqref{C12}. Any other NC inequality capable of revealing fully contextual quantum correlations requires a higher number of yes-no tests $\Pi_i$ or a higher number of experiments involving yes-no tests $\Pi_i$.

The graph in Fig.~\ref{Fig1} has $\alpha(G)=2$, $\vartheta(G)=\alpha^*(G)=\frac{5}{2}$, which are, respectively, the upper bounds for NCHV theories, QM, and GP theories in \eqref{C12}, and $\theta'(G)=5$, which is the number of experiments. Alternatively, the bound for NCHV theories can be calculated by generating the $2^{10}$ possible combinations of values $1$ or $0$ for the ten observables $\Pi_i$. In the proof of the next result we will see how to achieve the maximum quantum value of inequality \eqref{C12}.\hfill \endproof

%%%%%%%%%%%%%%%%%%%%%%%%%%%%%%%%%%%%%%%%%%%%%%%%%%%%%%%%%%%%%%%%%%%

{\em Theorem 2:} The minimum dimension of the quantum system needed to violate inequality \eqref{C12} is six.

{\em Proof. } The minimum dimension for the maximum quantum violation of inequality \eqref{C12} corresponds to the minimum dimension needed to have an orthogonal representation $\{|v_{i}\rangle\}_{i=0}^9$ of the graph in Fig.~\ref{Fig1} such that there is a state $|\psi\rangle$, such that $\sum_{i=0}^{9} |\langle\psi|v_{i}\rangle|^{2}= \frac{5}{2}$, which is $\vartheta(G)$ for the graph in Fig.~\ref{Fig1}. The minimum dimension needed to have an orthogonal representation of $G$ is called the orthogonal rank of $G$, $\xi(G)$. The graph in Fig.~\ref{Fig1} is the $J(5,2)$-Johnson graph which is the complement graph (i.e., the graph such that two vertices are adjacent if and only if they are not adjacent in the original graph) of the most famous ten-vertex graph, the Petersen graph; its orthogonal rank is $\xi(G)=6$.
%New in version 7
This can be proven by noticing that the orthogonality relations of the subgraph induced by vertices 0, 1, 2, 6, 7, and 9 cannot be implemented in dimension five, and giving an explicit orthogonal representation in dimension six of the graph in Fig.~\ref{Fig1}; for example, the following one:
\begin{subequations}
 \label{Vi}
\begin{align}
 \langle v_{0}| &= 8^{-1/2}(\sqrt{2},-\sqrt{2},0,0,2,0),\\
 \langle v_{1}| &= 8^{-1/2}(\sqrt{2},0,0,\sqrt{2},-1,\sqrt{3}),\\
 \langle v_{2}| &= 2^{-1}(1,-1,-1,-1,0,0),\\
 \langle v_{3}| &= 2^{-1}(1,-1,1,1,0,0),\\
 \langle v_{4}| &= 8^{-1/2}(\sqrt{2},0,0,-\sqrt{2},-1,\sqrt{3}),\\
 \langle v_{5}| &= 8^{-1/2}(\sqrt{2},0,-\sqrt{2},0,-1,-\sqrt{3}),\\
 \langle v_{6}| &= 8^{-1/2}(\sqrt{2},0,\sqrt{2},0,-1,-\sqrt{3}),\\
 \langle v_{7}| &= 2^{-1}(1,1,1,-1,0,0),\\
 \langle v_{8}| &= 8^{-1/2}(\sqrt{2},\sqrt{2},0,0,2,0),\\
 \langle v_{9}| &= 2^{-1}(1,1,-1,1,0,0).
\end{align}
\end{subequations}
This orthogonal representation is such that $\langle v_i | v_j \rangle=1$ if $i=j$, $\langle v_i | v_j \rangle=0$ if there is an edge between $i$ and $j$, and $\langle v_i | v_j \rangle=\frac{1}{2}$ otherwise. For this choice of yes-no tests $\Pi_i = |v_i\rangle \langle v_i|$, the following quantum state:
\begin{equation}
 \label{state}
 \langle\psi| =(1,0,0,0,0,0),
\end{equation}
is such that $\langle v_i | \psi \rangle=\frac{1}{2}$ for any $i=0,\ldots,9$ and, therefore,
\begin{equation}
 P(\Pi_i+\Pi_{i+1}+\Pi_{i+5}+\Pi_{i+7}=1)=1,
\end{equation}
for any $i=0,\ldots,4$. Therefore, state \eqref{state} provides the maximum quantum violation of inequality \eqref{C12}, proving that dimension six is also sufficient for its maximum quantum violation.\hfill \endproof

%%%%%%%%%%%%%%%%%%%%%%%%%%%%%%%%%%%%%%%%%%%%%%%%%%%%%%%%%%%%%%%%%%%

{\em Proposed experiments.} Fully contextual correlations for quantum systems of higher dimensionality than those of previous experiments \cite{ADLPBC12} can be observed by preparing the state \eqref{state} using, e.g., the polarization and three spatial modes of single photons, and then performing the five experiments needed for testing inequality \eqref{C12}. Each of these experiments consists of a sequential measurement of four compatible yes-no tests $\Pi_i=|v_i\rangle \langle v_i|$, where $|v_i\rangle$ is a unit vector belonging to the set \eqref{Vi} and represents a state of polarization and spatial modes of a single photon. The sequential measurements can be carried out by encoding the results of the previous measurements in time delays \cite{AACB12}, avoiding the complexity of previous schemes for sequential measurements on a single photon \cite{ADLPBC12,ARBC09}.

Another interesting experiment involving the state \eqref{state} and the projectors onto the states \eqref{Vi} is to show the genuinely six-dimensional impossible-to-beat quantum advantage for solving the task defined in Ref.~\cite{NDSC12} for the graph of Fig.~\ref{Fig1}. This experiment is feasible by exploiting the extra dimensions generated by combining the polarization and the orbital angular momentum of photons \cite{NSMSS10}.

%%%%%%%%%%%%%%%%%%%%%%%%%%%%%%%%%%%%%%%%%%%%%%%%%%%%%%%%%%%%%%%%%%%

{\em Conclusions.} We knew that five experiments involving yes-no tests (i.e., observables represented by rank-one projectors) are sufficient to reveal quantum contextual correlations on systems of dimension three \cite{KCBS08}. These experiments have been recently performed \cite{AACB12} (see also Ref.~\cite{LLSLRWZ11} for a related test involving six experiments), confirming the form of contextuality predicted by QM. However, for this scenario, more general theories can be more contextual than QM. Here we have shown that, surprisingly, five experiments are also sufficient to reveal quantum correlations with zero noncontextual content, which means correlations that are as contextual as they are allowed to be by the laws of probability; so even more general theories cannot be more contextual than QM. The price we have to pay for this impossible-to-beat quantum contextuality is using quantum systems of dimension six and longer sequences of yes-no tests.

The importance of the NC inequality \eqref{C12} goes beyond its simplicity; we have also shown that there is no scenario involving a smaller number of yes-no tests or a smaller number of experiments revealing fully contextual quantum correlations.

Moreover, we have found that the simplest NC inequality violated by QM, inequality \eqref{KCBS}, and the simplest NC inequality violated by QM as much as allowed by probability, the new inequality \eqref{C12}, are ``twin'' inequalities in many respects: the same number of experiments, the same pentagonal symmetry, the same upper bound for NCHV theories, the same upper bound for GP theories, the number of yes-no tests in \eqref{C12} is exactly double than that in \eqref{KCBS}, and the length of the sequences of compatible measurements in \eqref{C12} is exactly double than that in \eqref{KCBS}.

This similarity allows us to reformulate the question of what is the physical principle responsible of the peculiar form of quantum contextuality in a simple way: What physical principle limits quantum contextuality in the scenario of inequality \eqref{KCBS} but does not limit it in the (very similar) scenario of inequality \eqref{C12}?

%%%%%%%%%%%%%%%%%%%%%%%%%%%%%%%%%%%%%%%%%%%%%%%%%%%%%%%%%%%%%%%%%%%

{\em Note added.} After completing this manuscript we have found the answer to the last question: Surprisingly, the principle that limits quantum contextuality in the scenario of inequality \eqref{KCBS} is exactly the same principle that limits quantum contextuality in the scenario of inequality \eqref{C12}. This result is presented in \cite{Cabello12b}. The twin inequality shows explicitly something that is hidden in the KCBS inequality and only appears when two independent experiments testing the KCBS inequality are considered (see the details in Ref.~\cite{Cabello12b}): The graph representing this double KCBS experiment is a 25-vertex graph $G$ which has the same property of the (much simpler) ten-vertex graph of the twin inequality, namely, $\vartheta(G)=\alpha^*(G)$. The twin shows on a physical system of dimension 6 the same that the KCBS shows on a system of dimension 9. In a nutshell, the twin makes it easy to understand what is hard to understand in the KCBS inequality: the reason for its maximum quantum violation. One might argue that this changes the roles of the two inequalities: It is the twin which tells us what is physically relevant.

%%%%%%%%%%%%%%%%%%%%%%%%%%%%%%%%%%%%%%%%%%%%%%%%%%%%%%%%%%%%%%%%%%%

{\em Acknowledgment.} The author thanks C.~Budroni, L.~E.~Danielsen, A.~J.~L\'{o}pez-Tarrida, J.~R.~Portillo, S.~Severini, and A.~Winter for useful discussions, and C.B. for independently checking that inequality \eqref{C12} is tight. This work was supported by the Project No.\ FIS2011-29400 (Spain).

%%%%%%%%%%%%%%%%%%%%%%%%%%%%%%%%%%%%%%%%%%%%%%%%%%%%%%%%%%%%%%%%%%%

%%%%%%%%%%%%%%%%%%%%%%%%%%%%%%%%%%%%%%%%%%%%%%%%%%%%%%%%%%%%%%%%%%%


\begin{thebibliography}{99}

\bibitem{Specker60}
 E. P. Specker,
 %Die Logik nicht gleichzeitig entscheidbarer Aussagen.
 Dialectica \textbf{14}, 239 (1960).

\bibitem{Bell66}
 J. S. Bell,
 %On the problem of hidden variables in quantum mechanics.
 Rev. Mod. Phys. \textbf{38}, 447 (1966).

\bibitem{KS67}
 S. Kochen and E. P. Specker,
 %The problem of hidden variables in quantum mechanics.
 J. Math. Mech. \textbf{17}, 59 (1967).

\bibitem{CSW10}
 A. Cabello, S. Severini, and A. Winter,
 %(Non-)Contextuality of physical theories as an axiom.
 \eprint{arXiv:1010.2163}.

\bibitem{Fuchs11}
 C. Fuchs,
 %Some negative remarks on operational approaches to quantum theory.
 Perimeter Institute Recorded Seminar Archive 11050055,
 \url{http://pirsa.org/11050055/}

\bibitem{Cabello11}
 A. Cabello,
 %Correlations without parts.
 Nature (London) \textbf{474}, 456 (2011).

\bibitem{Cabello12}
 A. Cabello,
 in {\em A Computable Universe},
 edited by H. Zenil
 (World Scientific, Singapore, 2012), Chap.~31.

\bibitem{Cabello03}
 A. Cabello,
 %Kochen-Specker theorem for a single qubit using positive operator-valued measures.
 Phys. Rev. Lett. \textbf{90}, 190401 (2003).

\bibitem{Spekkens05}
 R. W. Spekkens,
 %Contextuality for preparations, transformations, and unsharp measurements.
 Phys. Rev. A \textbf{71}, 052108 (2005).

\bibitem{KCBS08}
 A. A. Klyachko, M. A. Can, S.~Binicio\u{g}lu, and A.~S.~Shumovsky,
 %Simple test for hidden variables in spin-1 systems.
 Phys.~Rev.~Lett. \textbf{101}, 020403 (2008).

\bibitem{ADLPBC12}
 E. Amselem, L. E. Danielsen, A.~J.~L\'{o}pez-Tarrida, J.~R.~Portillo, M.~Bourennane, and A.~Cabello,
 %Experimental fully contextual correlations.
 Phys.~Rev.~Lett. \textbf{108}, 200405 (2012).

\bibitem{Porta}
 \url{http://www.zib.de/Optimization/Software/porta/}

\bibitem{Diestel10}
 R. Diestel,
 {\em Graph Theory}, Graduate Texts in Mathematics, Vol.~173
 (Springer, Heidelberg, 2010).

\bibitem{Lovasz79}
 L. Lov\'{a}sz,
 %On the Shannon capacity of a graph.
 IEEE~Trans.~Inf.~Theory \textbf{25}, 1 (1979).

\bibitem{SU97}
 E. R. Scheinerman and D. H. Ullman,
 {\em Fractional Graph Theory}
 %Wiley-Interscience Series in Discrete Mathematics and Optimization
 (John Wiley \& Sons, New York, 1997).

\bibitem{CDLP11}
 A. Cabello, L. E. Danielsen, A.~J.~L\'{o}pez-Tarrida, and J.~R.~Portillo,
 \url{http://www.ii.uib.no/~larsed/quantum_graphs/}

\bibitem{AACB12}
 J. Ahrens,
 E. Amselem,
 A. Cabello, and
 M. Bourennane,
 \eprint{arXiv:1301.2887}.

\bibitem{ARBC09}
 E. Amselem, M. R{\aa }dmark, M.~Bourennane, and A.~Cabello,
 %State-independent quantum contextuality with single photons.
 Phys.~Rev.~Lett. \textbf{103}, 160405 (2009).

\bibitem{NDSC12}
 E. Nagali, V. D'Ambrosio, F.~Sciarrino, and A.~Cabello,
 %Experimental observation of impossible-to-beat quantum advantage on a hybrid photonic system.
 Phys.~Rev.~Lett. \textbf{108}, 090501 (2012).

\bibitem{NSMSS10}
 E. Nagali, L. Sansoni, L. Marrucci, E.~Santamato, and F.~Sciarrino,
 %Experimental generation and characterization of single-photon hybrid ququarts based on polarization and orbital angular momentum encoding.
 Phys.~Rev.~A \textbf{81}, 052317 (2010).

\bibitem{LLSLRWZ11}
 R. {\L}apkiewicz, P. Li, C. Schaeff, N. Langford, S.~Ramelow, M.~Wie\'{s}niak, and A.~Zeilinger,
 %Most basic experimental falsification of non-contextuality.
 Nature (London) \textbf{474}, 490 (2011).

\bibitem{Cabello12b}
 A. Cabello,
 %Simple explanation of the quantum violation of a fundamental inequality.
 Phys. Rev. Lett. \textbf{110}, 060402 (2013).

\end{thebibliography}
\end{document}